
\input epsf   

%
%
\global\newcount\meqno
\def\eqn#1#2{\xdef#1{(\secsym\the\meqno)}
\global\advance\meqno by1$$#2\eqno#1$$}
%
%
\global\newcount\refno
\def\ref#1{\xdef#1{[\the\refno]}
\global\advance\refno by1#1}
\global\refno = 1
\vsize=7.5in
\hsize=5in
\magnification=1200
\tolerance 10000
\font\sevenrm = cmr7

\hyphenation{non-over-lapping}

\def\lsim{ \,\, \vcenter{\hbox{$\buildrel{\displaystyle <}\over\sim$}}
 \,\,}

\def\s#1{{\bf#1}}
\def\sp{ \s p}
\def\sx{\s x}

\def\pmb#1{\setbox0=\hbox{$#1$}%
\kern-.025em\copy0\kern-\wd0
\kern.05em\copy0\kern-\wd0
\kern-.025em\raise.0433em\box0 }

\baselineskip 12pt plus 1pt minus 1pt
%
%

\medskip
\nobreak
\medskip

\vskip 2in
\medskip
\centerline{\bf OPTIMIZATION OF RENORMALIZATION GROUP FLOW}
\vskip 24pt
\centerline{Sen-Ben Liao$^{1,2}$\footnote{$^\dagger$}{electronic address:
senben@phy.ccu.edu.tw}, 
Janos Polonyi$^{1,3}$\footnote{$^\ddagger$}
{electronic address: polonyi@fresnel.u-strasbg.fr}
and Michael Strickland $^4$\footnote{$^*$}{electronic address:
strickland.41@osu.edu}}
\vskip 12pt
\centerline{\it Laboratory of Theoretical Physics $^{1}$}
\centerline{\it Louis Pasteur University}
\centerline{\it Strasbourg, France}
\vskip 12pt
\centerline{\it Department of Physics $^2$}
\centerline{\it National Chung-Cheng University}
\centerline{\it Chia-Yi, Taiwan R. O. C.}
\vskip 12pt
\centerline{\it Department of Atomic Physics $^3$}
\centerline{\it Lor\'and E\"otv\"os University}
\centerline{\it Budapest, Hungary}
\vskip 12pt
\centerline{\it and}
\vskip 12pt
\centerline{\it Department of Physics $^4$}
\centerline{\it Ohio State University }
\centerline{\it Columbus, Ohio\ \ 43210\ \ \ U.S.A.}
\vskip 1.2in
\centerline{Submitted to {\it Nucl. Phys. B}}
\vskip 24 pt
\baselineskip 12pt plus 2pt minus 2pt
\centerline{{\bf ABSTRACT}}
\medskip
\medskip

Renormalization group flow equations
for scalar $\lambda\Phi^4$ are generated using
three classes of smooth smearing functions.  Numerical results
for the critical exponent $\nu$ in three dimensions
are calculated by means of a truncated series expansion of the blocked
potential.
We demonstrate how the convergence of $\nu$ as a function of
the order of truncation can be
improved through a fine tuning of the smoothness
of the smearing functions. 

\vskip 24pt
\vfill
\noindent CCU-TH-99-01 \hfill May, 1999
\eject

\centerline{\bf I. INTRODUCTION}
\medskip
\nobreak
\xdef\secsym{1.}\global\meqno = 1
\medskip
\bigskip

Renormalization group (RG) methods provide a powerful tool 
for investigating nonperturbative physical phenomena \ref\wilson. 
Issues such as QCD under extreme conditions,
formation of a quark-gluon plasma in relativistic heavy-ion collisions
\ref\polonyi, or
critical phenomena in condensed matter systems all cannot be treated
using standard perturbation theory due to the presence of
infrared (IR) singularities. Through the continuous elimination of degrees
of freedom, RG techniques systematically resum the perturbative series and 
therefore can provide information about nonperturbative effects in
these systems.  However, the power of RG relies on the existence
of efficient analytic and computational methods since the full RG flow
equations cannot be solved exactly.  The goal of this
paper is to show how to optimize the way the degrees of freedom
are eliminated in order to improve results obtained using approximate RG
flow equations.

For a field theoretical system a continuous RG transformation can be realized
by introducing a smearing
function $\rho_k(x)$ which governs the coarse-graining procedure 
\ref\lp \ref\sb. The
scale $k$ acts as an effective IR cutoff that
separates the low- and
high-momentum modes. Using this smearing function an 
averaged blocked field can be defined as
\eqn\bloc{\phi_k(x)=\int_y\rho_k(x-y)\phi(y),}
from which one obtains the effective Legendre blocked action: 
\eqn\cactt{ e^{-\widetilde S_k[\Phi(x)]}=\int
D[\phi]\prod_{x}
\delta(\phi_k(x)-\Phi(\sx))e^{-S[\phi]}\,.}
In this manner we achieve a smooth interpolation between the bare action 
$S_{\Lambda}[\Phi]$ defined at the ultraviolet (UV) cutoff, $\Lambda$,
and the quantum effective action $\widetilde S_k[\Phi]$, 
%
%
which generates those one-particle-irreducible graphs 
whose internal momenta extend between $k$ and $\Lambda$ 
\ref\tmorris.
The introduction of a smearing function leads to a modification of the
bare propagator:
\eqn\smprop{ \Delta(p)={1\over p^2}\longrightarrow \Delta_k(p)
={{1-\rho_k(p)}\over p^2}={{\tilde\rho_k(p)}\over p^2},}
with $\rho_k(p)+\tilde\rho_k(p)=1$. One can then derive
a RG equation for the blocked action which remains
valid in all orders of the loop expansion by varying ${\widetilde S}_k$
infinitesimally with $k$:
\eqn\exac{\eqalign{ k{{\partial {\widetilde S}_k} \over {\partial k}} &=
-{1\over 2}{\rm Tr}\Biggl[{1\over \Delta_k}\Bigl(k{\partial\Delta_k\over
{\partial k}}\Bigr)\biggl(1+\Delta_k~{\delta^2{\widetilde S}_k\over
{\delta\Phi^2}}\biggr)^{-1}\Biggr] \cr
&
=-{1\over 2}{\rm Tr}\Biggl[{\tilde\rho_k}^{-1}
\Bigl(k{\partial{\tilde\rho_k}\over{\partial k}}
\Bigr) \Bigl(1+{{\tilde\rho_k}\over p^2}~
{\delta^2{\widetilde S}_k\over{\delta\Phi^2}}\Bigr)^{-1}
\Biggr].}}
Since the pioneering work of Wilson \wilson, similar RG equations 
have been derived and analyzed. A sharp momentum cutoff was first used by
Wegner and Houghton \ref\wegner, and further discussed in Refs. 
\ref\hasenfratz, \ref\margaritis\ and \ref\aoki. Polchinski 
\ref\polchinski, on the other hand, employed a smooth cutoff. 
There exists a vast amount of literature devoted to this subject \ref\others.
Such functional RG equations, as stated before, 
have proven too difficult to be solved 
exactly, and further approximations are needed.
Any viable scheme must not only retain the nonperturbative 
characteristics of the RG, but also converge sufficiently rapidly without
inducing further spurious effects. 
As demonstrated by Morris in his
seminal papers \tmorris \ref\morris, the most reliable method so far 
for probing the low-energy effective theory is the
derivative expansion:
\eqn\deri{ {\widetilde S}_k[\Phi]=\int_x\biggl\{ {Z_k(\Phi)\over 2}
(\partial_{\mu}\Phi)^2+U_k(\Phi)+O(\partial^4)\biggr\},}
where $Z_k(\Phi)$ and $U_k(\Phi)$ are, respectively, the wavefunction
renormalization and the blocked potential. 
At leading order in
the derivative expansion $Z_k(\Phi)$ is taken to be unity and the 
low-energy effective action is described solely in terms
of $U_k(\Phi)$ \ref\nicoll. This local potential approximation (LPA) 
results in the following flow equation:
\eqn\rg{ k{{\partial U_k(\Phi)} \over {\partial k}}=
{1\over 2}\int_p \Bigl(k{{\partial\tilde\rho_k(p)}\over {\partial k}}
\Bigr)~{U_k''(\Phi)\over{ p^2+\tilde\rho_k(p)
U''_k(\Phi) }}.}
Although the above expression can be derived exactly \ref\jens,
it may also be obtained by a differentiation of 
the perturbative one-loop result
\eqn\oneloop{ {\tilde U}^{(1)}_k(\Phi)={1\over 2}\int_p{\rm ln}~
\biggl\{1+\tilde\rho_k(p){V''(\Phi)\over p^2}\biggr\}}
with respect to ${\rm ln}(k)$ followed by
a substitution $V''(\Phi) \to U''_k(\Phi)$ on the right-hand-side.

Clearly, the functional form of the RG equation for $U_k(\Phi)$ 
depends on the choice of the smearing function. 
A sharp cutoff, $\rho_k(p)=\Theta(k-p)$, 
provides a well-defined boundary between the high-
and low-momentum modes and yields a nonlinear 
partial differential equation. On the other hand, for 
a general smooth cutoff, no clear separation
exists and the RG flow remains an integro-differential equation.
According to the universality principle the shape of the smearing
function, $\rho_k(p)$, does not influence the physics
of finite length scales. So long as the
effective action contains all marginal or relevant
operators the resulting RG flow must
be scheme independent.
At leading order in the derivative expansion, 
with no further approximation involved, 
physical results are indeed independent of the shape of $\rho_k(p)$,
as demonstrated in \tmorris\ref\ball. 
However, at next-to-leading order in the derivative expansion
where the effect of $Z_k(\Phi)$ is included, 
the critical properties of the system will depend on  
whether $\rho_k(p)$ is sharp or smooth. In fact, ambiguities arise
in the former, although self-consistency arguments can be used
to circumvent this difficulty \ref\mike. The dependence 
of $\eta$, the anomalous dimension, on the shape of the smearing function
beyond leading order, has been calculated in  
\ball\ref\sbl.

The critical exponents and other universal properties of the
system may vary when they are determined in the vicinity of 
the fixed point of the truncated equation given in Eq. \rg, or
when the theory moves far away from the critical manifold. In fact,
even if the solution includes all the relevant operators,
the neglected irrelevant parameters are still evolving
at the fixed point of the truncated equation. Since
the modification of the irrelevant operators amounts to
a change of an overall scale factor of the theory this
scale factor remains cutoff dependent at the approximate fixed
point. In order to minimize its variation one has to 
retain in the approximation the irrelevant operators
with the critical exponent.

A frequently used technique is to
expand $U_k(\Phi)$ in power series of $\Phi$ followed by a truncation
at some order, thereby turning the problem into solving a set of 
coupled nonlinear ordinary differential equations. 
Although the numerical algorithms are simplified by this expansion, the
convergence of critical exponents calculated in this way is not guaranteed.  
In fact, when employing a sharp smearing function while $U_k(\Phi)$ 
is expanded in $\Phi$ about the origin, the critical exponent $\nu$
has been shown to oscillate
about its expected value as $M$, the number of terms in the series, is 
increased \morris. The convergence improves, however, if
the expansion is made around the $k$-dependent
minimum of $U_k(\Phi)$ \aoki. Unfortunately, this approach fails 
when applied to an $O(N)$ field theory in the symmetry-broken phase 
due to the presence of the massless Goldstone
modes which cause IR divergences to persist 
\jens \ref\bergerhoff.  It is therefore desirable to identify a computational
method which does not suffer from these drawbacks.
In addition, for gauge theories, the numerical complication of solving the
non-truncated flow equations is even more
overwhelming because of the proliferation of 
degrees of freedom and the complexity
of their interactions. For these theories a polynomial truncation of 
the blocked potential seems inevitable. Thus,
it is crucial to understand how to
improve results obtained from truncation schemes.  

In the present work, we examine what happens, in the truncated
polynomial expansion of $U_k(\Phi)$, when a smooth smearing function
$\rho_{k,\sigma}(p)$ with $\sigma$ being the smoothness parameter, is 
utilized instead of $\Theta(k-p)$. In particular, we explore how
$\nu$ changes with $M$ as the
smooth smearing functions approach the sharp limit, and how its convergence
is influenced by $\sigma$.
Our goal is to obtain a prescription which eliminates or 
diminishes the oscillations mentioned
above and ensures a rapid convergence as $M$ is increased.

The organization of the paper is as follows. In Sec. II we give three
examples of smooth smearing functions and derive
the corresponding RG equations.
We show how the expected sharp cutoff limit can be recovered by
recognizing that $\rho_k(p)$ is a {\it continuous} function in the 
vicinity   
$p\approx k$. Finite-temperature RG equations are also derived. 
In Sec. III we present the numerical solutions for 
$\nu$ associated with the three-dimensional Wilson-Fisher fixed point 
at various levels of polynomial truncation of $U_k(\Phi)$. In Sec. IV 
the source of scheme dependence on $M$ and $\sigma$ is discussed. We 
propose an optimized smooth smearing function which 
leads to a maximal cancellation of the effects of irrelevant operators and
provides 
the fastest convergence of $\nu$ to a value which is in good agreement
with the world's best estimate. 
Sec. V is reserved for summary and discussions.

\bigskip
\medskip

\centerline{\bf II. SMOOTH CUTOFF FUNCTIONS}
\medskip
\nobreak
\xdef\secsym{2.}\global\meqno = 1
\medskip
\nobreak

We begin by considering a general smooth representation of the 
smearing function $\rho_{k,\varepsilon}(p)=
\Theta_{\varepsilon}(k,p)$ which approaches $\Theta(k-p)$ as 
$\varepsilon\to 0$. Due to the singular nature of the 
step function ambiguities can arise 
when taking the limit $\varepsilon\to 0$.  More precisely,
one will encounter terms involving $\Theta(0) \equiv \theta_0$.
Since this value
is not uniquely specified for the smooth parameterizations of the step 
function,
it would seem that the approach to the sharp limit is not unique.  
The naive, or {\it mean}, approach in the limit $\varepsilon\to 0$ is:
\eqn\lemm{ \int_0^{\infty}dp~
{\partial\Theta_{\varepsilon}(k,p)\over{\partial k}}
G(\Theta_{\varepsilon}(k,p),p)\longrightarrow \int_0^{\infty}dp~\delta(k-p)
G(\theta_0,p)=G(\theta_0,k),}
where $\partial_k\Theta_{\varepsilon}(k,p)\to \delta(k-p)$.
This prescription, however, is oversimplified 
because the integrand of
\eqn\nuint{ \int_0^{\infty}dp~G(\Theta_{\varepsilon}(k,p),p)),}
is not uniformly convergent as $\varepsilon\to0$. The proper
treatment requires that $\Theta_{\varepsilon}(k,p)$ be
used as an integration variable. Thus by setting
$\Theta_{\varepsilon}(k,p)=\Theta_{\varepsilon}(1-p/k)=t$ 
and making a change of variables from
$p$ to $t$, we have 
\eqn\lemn{\int_0^{\infty}dp~
{\partial\Theta_{\varepsilon}(k,p)\over{\partial k}}
G(\Theta_{\varepsilon}(k,p),p)  \longrightarrow \int_0^{\infty}dp\bigl(
-{p\over k}{dt\over dp}\bigr)G(t,p)=-\int_{t(p=0)}^{t(p=\infty)}dt~
{p(t)\over k}~G(t,p(t)).}
The sharp cutoff limit is obtained when $t(p=0)=1$, $t(p=\infty)=0$, 
$p(t)=k$ and Eq. \lemn\ simplifies to 
\eqn\lesh{ \int_0^{\infty}dp~
{\partial\Theta_{\varepsilon}(k,p)\over{\partial k}}
G(\Theta_{\varepsilon}(k,p),p)=\int_0^1dt~G(t,k).} 
In terms of $t$, the RG flow equation in Eq. \rg\ reduces to
\eqn\rgt{\eqalign{
k{{\partial U_k(\Phi)} \over {\partial k}} &=
{S_d\over 2}\int_{t(p=0)}^{t(p=\infty)}dt~
{p^d(t)~U_k''(\Phi)\over{ p^2(t)+(1-t)U''_k(\Phi)}} \cr
&
={S_dk^d\over 2}\int_{t(z=0)}^{t(z=\infty)}dt~
{z^d(t)~{\bar U}_k''(\Phi)\over{ z^2(t)+(1-t){\bar U}''_k(\Phi)}},}}
where $S_d=2/(4\pi)^{d/2}\Gamma(d/2)$, 
$z(t)=p(t)/k$ and ${\bar U}_k''(\Phi)=U_k''(\Phi)/k^2$.
Thus, with a given $\rho_{k,\varepsilon}(p)\equiv t$, we first invert
the expression to obtain $z(t)$ and then make a substitution into Eq. \rgt\ to
deduce the corresponding smooth RG equation for $U_k(\Phi)$.

As stated in the Introduction, when a truncated polynomial expansion is 
used, physical quantities such as
the critical exponents may depend on $M$ as well as on $\sigma$.
For a sharp cutoff, the critical exponent $\nu$ has been
shown to exhibit oscillatory behavior 
about its expected value when $U_k(\Phi)$ is expanded about
the origin or the $k$-dependent minimum \hasenfratz \morris, 
although the result improves significantly in the latter \aoki. 
In order to determine whether these oscillations can be removed or reduced
by using a smooth smearing function
we consider below three parameterizations of $\rho_{k,\sigma}(p)$, all of
which approach $\Theta(k-p)$, when the appropriate
limit of the smoothness parameter $\sigma$ is taken.

\medskip
\bigskip

\eject

\centerline{\bf A. Hyperbolic Tangent}
\medskip

\noindent
We first examine
\eqn\smooth{ \rho_{k,\varepsilon}(p)=
{1\over 2}\Bigl[1+{\rm tanh}\Bigl({{k^2-p^2}\over{p k \varepsilon}}\Bigr)
\Bigr].}
The above smearing function satisfies:
\medskip
\eqn\proper{\eqalign{ ({\rm i}) &\quad \lim_{\varepsilon\to 0}\rho_{k,
\varepsilon}(p) = \Theta(k-p),\cr
({\rm ii}) &\quad \lim_{k\to 0} \rho_{k,\varepsilon}(p) = {1\over 2}\bigl[
1+{\rm tanh}(-\infty)\bigr]=0,\cr
({\rm iii}) &\quad \lim_{k \to \infty} \rho_{k,\varepsilon}(p) = 1.}}
In this case the propagator is modified as:
\eqn\smprop{ \Delta_{k,\varepsilon}(p)={1\over p^2}\cdot{1\over 2}
\Bigl[1+{\rm tanh}\Bigl({{p^2-k^2}\over{p k \varepsilon}}\Bigr)\Bigr].}
Substituting
\eqn\disme{ k{{\partial\rho_{k,\varepsilon}}\over {\partial k}}
={ k^2 + p^2 \over 2 k p \varepsilon}~ 
{\rm sech}^2\Bigl({k^2 - p^2 \over k p \varepsilon}\Bigr),}
into Eq. \rg\ then yields:
\eqn\smrg{ k{{\partial U_k(\Phi)} \over {\partial k}}=
-{S_d\over{4 \varepsilon}}\int_0^{\infty}dp~p^{d-1} 
\Bigl({ k\over p} + {p \over k}\Bigr)~
{\rm sech}^2\Bigl({k^2 - p^2 \over k p \varepsilon}\Bigr)
{U_k''(\Phi)\over{ p^2+\bigl[1-\rho_{k,\varepsilon}(p)\bigr]U''_k(\Phi) }}.}
With the help of Eq. \lesh\ we can rewrite the above expression as
\eqn\rgt{
k{{\partial U_k(\Phi)} \over {\partial k}}=
{S_dk^d\over 2}\int_{t(z=0)}^{t(z=\infty)}dt~
{z^d(t)~{\bar U}_k''(\Phi)\over{ z^2(t)+(1-t){\bar U}''_k(\Phi)}},}
where 
\eqn\pt{\eqalign{ z(t)&=\sqrt{ {\varepsilon^2 a^2(t)\over 4} + 1 } 
- {\varepsilon a(t) \over 2} \cr
	  a(t) &= {\rm tanh}^{-1}( 2 t -1 ) .}}
One of our main goals here is to examine how
the flow of the theory is influenced by tuning $\varepsilon$. 
As $\varepsilon\to 0$, the expected sharp-cutoff Wegner-Houghton 
equation \wegner\
\eqn\scrg{ k{{\partial U_k(\Phi)} \over {\partial k}}=
-{S_dk^d\over 2}\int_0^1dt~{{\bar U}''_k(\Phi)\over{1
+(1-t){\bar U}''_k(\Phi)}}
=-{S_dk^d\over 2}{\rm ln}\biggl[{{k^2+U''_k(\Phi)}\over k^2}\biggr]}
is recovered. As $\varepsilon$ is increased the cutoff becomes smoother,
i.e., the peak in $\partial\rho_{k,\varepsilon}(p)/\partial p$
at $p=k$ spreads.
At about $\varepsilon\approx 2$ this tendency is reversed and
the peak becomes taller again. The location of the peak stays roughly
the same until $\varepsilon\approx1$ and is shifted towards
smaller values as $\varepsilon$ is further increased.
It finally moves to $p=k/\varepsilon$ for large $\varepsilon$.
The decrease of the cutoff when $\varepsilon\to\infty$
eliminates the interactions and the  
theory flows into the trivial Gaussian fixed point. We shall comment
more on this point later. 

Notice that the mean approach described in Eq. \lemm\ gives
\eqn\smrf{\eqalign{ k{{\partial U_k(\Phi)} \over {\partial k}}
&=\lim_{\varepsilon\to 0}~
-{S_d\over{4 \varepsilon}}\int_0^{\infty}dp~p^{d-1} 
\Bigl({k\over p}+{p\over k}\Bigr)~{\rm sech}^2\Bigl({k^2-p^2\over kp
\varepsilon}\Bigr){U_k''(\Phi)\over{p^2+\bigl[1-\rho_{k,\varepsilon}(p)\bigr]
U''_k(\Phi)}} \cr
&
\longrightarrow -{S_d\over 2}\int_0^{\infty}dp~p^{d-1}~k\delta(k-p)~
{U''_k(\Phi)\over{p^2+\theta_0U''_k(\Phi)}}
=-{S_dk^d\over 2}~\biggl({U_k''(\Phi)\over{k^2+\theta_0
U_k''(\Phi)}}\biggr),}}
which differs from the Wegner-Houghton equation in its absence of the
characteristic logarithmic functonal structure. 
In fact, the RG equation is highly sensitive to the $k$ dependence in
$\rho_{k,\sigma}(p)$. Nevertheless, we shall see that the difference
between the two equations does not 
affect the critical properties significantly in the next section.  

Next, we turn the nonlinear 
flow of $U_k(\Phi)$ into a set of coupled ordinary differential 
equations by making an expansion in power series of $\Phi$:
\eqn\uexpa{ U_k(\Phi)=\sum_{\ell=1}^{\infty}{
g^{(2\ell)}_k\over{(2\ell)!}}~\Phi^{2\ell},\qquad\qquad 
g^{(2\ell)}_k=U^{(2\ell)}_k(0)={{\partial^{2\ell} U_k}\over{\partial
\Phi^{2\ell}}}\Big\vert_{\Phi=0},}
followed by a truncation at some order $\ell=M$. 
The resulting running equations for the first two terms read: 
\eqn\nwo{\eqalign{k{\partial{\bar\mu}_k^2\over{\partial k}}&=-2{\bar\mu}^2_k
+J_0{\bar\lambda}_k,\cr
k{\partial{\bar\lambda}_k\over{\partial k}}&=-\epsilon{\bar\lambda}_k
-{{6\bar\lambda}_k^2\over{\bar\mu_k^2}}\bigl(J_0-J_1\bigr)
+J_0{\bar g}^{(6)}_k,}}
where $\epsilon=4-d$,  
\eqn\defi{ J_n(\bar\mu_k^2)=-{S_d\over 2}\int_0^1
dt~{z(t)^{d+2(n+1)}\over
{\bigl[z^2(t)+(1-t)\bar\mu_k^2\bigr]^{2+n}}},}
and $z(t)$ is given by Eq. \pt. 
In the sharp cutoff limit where $z(t)\to 1$ and  
\eqn\zts{ J_n(\bar\mu_k^2)={S_d\over{2(1+n)}}~{{1-(1+\bar\mu_k^2)^{1+n}}\over
{\bar\mu_k^2(1+\bar\mu_k^2)^{1+n}}},}
we recover
\eqn\shwo{\eqalign{k{\partial{\bar\mu}_k^2\over{\partial k}}&=
-2\bar\mu_k^2-{S_d\over 2}
{\bar\lambda_k\over{1+\bar\mu_k^2}},\cr
k{\partial{\bar\lambda}_k\over{\partial k}}&=
-\epsilon\bar\lambda_k+{S_d\over 2}\biggl\{
{3\bar\lambda_k^2\over{\bigl(1+\bar\mu_k^2\bigr)^2}}-{{\bar g}^{(6)}_k
\over{1+\bar\mu_k^2}}
\biggr\}.}}
This can be compared with the result obtained using the mean approach:
\eqn\nwo{\eqalign{k{\partial{\bar\mu}_k^2\over{\partial k}}&=
-2\bar\mu_k^2-{S_d\over 2}
{\bar\lambda_k\over{\bigl(1+\theta_0\bar\mu_k^2\bigr)^2}},\cr
k{\partial{\bar\lambda}_k\over{\partial k}}&=
-\epsilon\bar\lambda_k+{S_d\over 2}\biggl\{
{6\theta_0\bar\lambda_k^2\over{\bigl(1+\theta_0\bar\mu_k^2\bigr)^3}}
-{{\bar g}^{(6)}_k\over{\bigl(1+\theta_0\bar\mu_k^2\bigr)^2}}\biggr\}.}}
A comparison of the two flow equations shows that
in order to recover the standard perturbative RG coefficient functions
in the UV limit $\bar\mu_k^2 \ll 1$, 
the usual convention $\theta_0=1/2$ must be imposed. However,
this choice also implies a deviation of scaling 
for a massive theory in the IR 
regime where $\bar\mu^2_k \gg 1$.

It is easy to understand that the two flows agree
in the UV scaling regime only. In fact, the non-uniform
convergence in Eq. \nuint\ leads to an ``incorrect'' use of the 
integrand of the evolution equation Eq. \rg\ at $p\approx k$
when the limit $\varepsilon\to0$ is taken before the integration.
Since the integrand is independent of the field $\Phi$
for $k^2\gg U''_k(\Phi)$ the mistake is unimportant. 
But the naive evolution equation displays wrong $U''_k(\Phi)$ 
dependence when $k^2\approx U''_k(\Phi)$.

\bigskip
\medskip

\centerline{\bf B. Exponential Function}
\medskip

\noindent
An alternative class of smearing functions is the exponential function
\eqn\bsm{ \rho_{k,b}(p)=e^{-a(p/k)^b},}
where $a$ and $b$ are constants. This smearing function satisfies 
the following conditions:
\eqn\proper{\eqalign{({\rm i}) &\quad \lim_{b\to\infty}\rho_{k,b}(p)
=\Theta(k-p),\cr
({\rm ii}) &\quad \lim_{k\to 0} \rho_{k,b}(p)=0, \cr
({\rm iii}) &\quad \lim_{k\to \infty} \rho_{k,b}(p) = 1.}}
In order to facilitate comparison with the other smearing functions used here,
we also
require $\rho_{k,b}(p=k)
=1/2$, which in turn implies $a={\rm ln}~2$, or $\rho_{k,b}(p)=2^{-(p/k)^b}$. 

The corresponding RG equation reads
\eqn\brg{ 
k{{\partial U_k(\Phi)} \over {\partial k}}=
-{S_dk^d\over 2}~ab\int_0^{\infty}dy~y^{d+b-1}~{e^{-ay^b}
{\bar U}_k^{''}(\Phi)\over
{y^2+\bigl(1-e^{-ay^b}\bigr){\bar U}_k^{''}(\Phi)}}.}
Alternatively, one can also cast Eq. \brg\ into the same form as 
Eq. \rgt, but with $z(t)$ given by 
\eqn\zzt{ z(t)=\Bigl(-{{\rm ln}~t \over a}\Bigr)^{1/b}.}
In the asymptotic limit where
${\bar U}''_k(\Phi) \ll 1$, the RG flow can be approximated as
\eqn\rgapp{ k{{\partial U_k} \over {\partial k}}
\approx -{S_dk^d\over 2}\biggl[c_1~{U''_k\over k^2}
+c_2{{U_k''}^2\over k^4}+   O({U_k^{''3}\over k^6})\biggr],}
where
\eqn\amm{\eqalign{ c_1 &= 
a^{-(d-2)/b}~\Gamma\bigl({{b+d-2}\over b}\bigr),\cr
c_2 &=-\bigl[1-2^{-(d+b-4)/b}\bigr]a^{-(d-4)/b}~\Gamma
\bigl({{b+d-4}\over b}\bigr).}}
The sharp limit, on the other hand, gives $c_1=1$ and $c_2=-1/2$
independent of dimensionality $d$. Thus, $c_1$ and $c_2$ provide a 
measure of the deviation from the sharp-cutoff UV scaling. 
We also note that in order to avoid UV singularities in the
integrations, $b$ must be chosen as to avoid singularities in the Gamma functions
above. As $b$ is decreased the cutoff becomes smoother. The location
of the peak in $\partial\rho_{k,b}(p)/\partial p$ is stable
for large $b$ and starts to shift at $b\approx3$. For $b \lsim 3$,
the shape of $\rho_{k,b}(p)$ spreads in such a manner that
one can no longer associate a well-defined cutoff value.

\medskip
\bigskip

\centerline{\bf C. Power-law Function}
\medskip

\noindent
Lastly, we consider a power-law smearing function:
\eqn\alge{ \rho_{k,m}(p)={1\over{1+(p/k)^m}},}
which satisfies
\eqn\proper{\eqalign{({\rm i}) &\quad \lim_{m\to\infty}\rho_{k,m}(p)
=\Theta(k-p),~~[~\theta_0={1/2}~~~{\rm when}~p=k~],\cr
({\rm ii}) &\quad \lim_{k\to 0}\rho_{k,m}(p)=0, \cr
({\rm iii}) &\quad \lim_{k\to \infty}\rho_{k,m}(p)=1.}}
The RG equation in this case reads
\eqn\arg{\eqalign{ k{{\partial U_k(\Phi)} \over {\partial k}} &=
-{S_dk^d\over 2}~m\int_0^{\infty}dy~y^{d+m-1}~
{{\bar U}_k''(\Phi)\over{ \bigl(1+y^m\bigr)\bigl[y^2(1+y^m)+y^m 
~{\bar U}''_k(\Phi)\bigr]}} \cr
&
=-{S_dk^d\over 2}\int_0^{1}dt~
{z^d(t)~{\bar U}_k''(\Phi)\over{ z^2(t)+(1-t){\bar U}''_k(\Phi)}},}}
where $y=p/k$ and 
\eqn\ptt{ z(t)=\Bigl({1\over t}-1\Bigr)^{1/m}.}
The sharp limit, i.e., $z\to 1$, may also be obtained by the substitution:
\eqn\lem{ {my^{m-1}\over{\bigl(1+y^m\bigr)^2}}~G\bigl(
{1\over{1+y^m}},p\bigr)
\longrightarrow \delta(1-y)\int_0^1dt~G(t,p),\quad m\to\infty,}
which results from 
\eqn\sham{ \lim_{m\to\infty}~{my^{m-1}\over{(1+y^m)^2}}=\delta(1-y).}
The corresponding
flow equations for $\bar\mu_k^2$ and $\bar\lambda_k$ have the same
functional forms as those in Eq. \nwo\ but with
\eqn\defii{ J_n(\bar\mu_k^2)=
-{S_d\over 2}~m\int_0^{\infty}dz~{{(1+z^m)^nz^{d+2n+m+1}}\over
{\bigl[z^2(1+z^m)+z^m\bar\mu_k^2\bigr]^{2+n}}}.}
On the other hand, if one adopts the mean approach and takes 
the limit $m\to\infty$, the RG equation then becomes
\eqn\argu{ k{{\partial U_k(\Phi)} \over {\partial k}}\to
-{S_dk^d\over 2}\int_0^{\infty}dy~\delta(1-y)
{y^dU_k''(\Phi)\over{ y^2+\Theta(y-1)U''_k(\Phi)}}=-{S_dk^d\over 2}
{U''_k(\Phi)\over {k^2+\theta_0U''_k(\Phi)}},}
which again coincides with Eq. \smrf. The peak in
$\partial\rho_{k,m}(p)/\partial p$ is stable at $p=k$ for $m>5$
but shifts towards lower momenta for $m<5$, and 
the shape of the smearing function makes it difficult
to identify a clear cutoff value.

Unlike the previous two parameterizations where the integro-differential
flow equations must be solved numerically, for certain values of $d$ and $m$
analytic evaluation of the integral on the right-hand-side of Eq. \arg\ 
is possible.  For example, taking $d=m=4$ we have 
\eqn\rgf{ k{{\partial U_k} \over {\partial k}}=-{k^4\over 16\pi^2}
{U''_k\over{\sqrt{4k^4-{U''_k}^2}}}\biggl\{\pi-2~{\rm tan}^{-1}
\Bigl( {U''_k\over{\sqrt{4k^4-{U''_k}^2}}}\Bigr)\biggr\}.}
And for $d=3$, we have
\eqn\rgfr{ k{{\partial U_k}(\Phi) \over {\partial k}}=\cases{\eqalign{
& -{k^3\over 4\pi}\Bigl[-1+\sqrt{1+U''_k(\Phi)/k^2}~\Bigr],
\quad\qquad\qquad\qquad\quad\quad\qquad~~ (m=2), \cr
\cr
& -{k^3\over 4\pi}{1\over{\sqrt{2+U''_k(\Phi)/k^2}}}
\Bigl[-2+\sqrt{4+2U''_k(\Phi)/k^2}~\Bigr],\quad\qquad\quad (m=4).}}} 
In this differential form it is possible to apply the standard techniques 
of differential equations to solve for the RG flow of the theory.

In the asymptotic limit where ${\bar U}''_k(\Phi) \ll 1$, Eq. \arg\ becomes 
\eqn\apr{ k{{\partial U_k} \over {\partial k}}
\approx -{S_dk^d\over 2}\biggl[{\tilde c}_1~{U''_k\over k^2}
+{\tilde c}_2{{U_k''}^2\over k^4}+   O({{U''}_k^3\over k^6})\biggr],}
where
\eqn\amt{\eqalign{ {\tilde c}_1 &={m\over{d+m-2}}
\Gamma\bigl({{m-d+2}\over m}\bigr)
\Gamma\bigl({2m+d-2\over m}\bigr)~~
 {\buildrel {m\to\infty}\over\longrightarrow} ~~1,\cr 
{\tilde c}_2 &=-{m\over 2(d+2m-4)}\Gamma\bigl({{m-d+4}\over m}\bigr)
\Gamma\bigl({3m-d+4\over m}\bigr)~~
{\buildrel {m\to\infty}\over\longrightarrow} ~-{1\over 2}.}}

Before closing this section, we remark
that the RG techniques discussed so far
can be readily extended to finite-temperature systems.
For the scalar $\lambda\Phi^4$ theory defined on $S^1\times R^d$ 
in the imaginary-time formalism, with the radius of $S^1$ being given by 
$\beta$, the inverse temperature, one obtains 
\eqn\rgft{\eqalign{ k{{\partial U}_{\beta, k}(\Phi)\over {\partial k}} &=
-{S_dk^d\over 2\beta}\Biggl\{\beta\sqrt{k^2+U''_{\beta,k}(\Phi)}
+2~{\rm ln}\Bigl[1-e^{-\beta\sqrt{k^2+U''_{\beta,k}(\Phi)}}\Bigr]
\Biggr\} \cr
&
=-{S_dk^d\over \beta}~{\rm ln~sinh}\biggl({{\beta\sqrt{k^2+U''_{\beta,k}
(\Phi)}}\over 2}\biggr)\,}}
using a sharp cutoff \mike.
A smooth smearing function, on the other hand, 
yields an integro-differential equation:
\eqn\rgts{\eqalign{ k{{\partial U}_{\beta, k}(\Phi)\over {\partial k}} &=
-{1\over 4}\int_{\sp}
\Bigl(k{{\partial\rho_{k,\varepsilon}}\over {\partial k}}
\Bigr){U''_{\beta,k}\over{\sqrt{\sp^2+\bigl(1-\rho_{k,\varepsilon}\bigr)
U''_{\beta,k}}}}~{\rm coth}\Bigl({\beta\sqrt{\sp^2
+(1-\rho_{k,\varepsilon})U''_{\beta,k}}\over 2}\Bigr) \cr
&
\longrightarrow {S_dk^d\over 4}\int_{t(z=0)}^{t(z=\infty)}dt~
{z^d(t){\bar U}''_{\beta,k}\over{\sqrt{z^2(t)+(1-t)
{\bar U}''_{\beta,k}}}}~{\rm coth}\Bigl({\bar\beta\sqrt{z^2(t)
+(1-t){\bar U}''_{\beta,k}}\over 2}\Bigr),}}
where $\bar\beta=\beta k$. The two equations coincide
in the sharp cutoff limit.
In the high-temperature regime where $\bar\beta\to 0$, we have
\eqn\trg{ k{{\partial U}_{\beta, k}(\Phi)\over {\partial k}} \approx
{S_dk^d\over 2}\int_{t(z=0)}^{t(z=\infty)}dt~z^d(t)
\Biggl[{{\bar U}''_{\beta,k}(\Phi)\over{\bar\beta\bigl[z^2(t)
+(1-t){\bar U}''_k(\Phi)\bigr]}}+{\bar\beta\over 12}{\bar U}_{\beta,k}''(\Phi)
+O(\bar\beta^3)\Biggr].}
Apart from the factor of $\bar\beta^{-1}$, the first term in the bracket
corresponds to the zero-temperature $d$-dimensional case, in
accord with the expectation of dimensional reduction. In this limit, 
$\bar\beta$ can be scaled away with a redefinition of $U_{\beta,k}(\Phi)\to
\beta^{-1}U_k(\Phi)$ and $\Phi\to \beta^{-1/2}\Phi$ \mike \ref\cons.

\bigskip
\medskip

\centerline{\bf III. FIXED-POINT SOLUTIONS AND NUMERICAL RESULTS}
\medskip
\nobreak
\xdef\secsym{3.}\global\meqno = 1
\medskip
\nobreak

As seen in the last section, when the smearing function is smooth 
the flow of the theory is generally characterized by an
integro-differential equation, as opposed to
the differential equation obtained with the sharp cutoff or special cases
of a smooth cutoff. In dimensionless form, the flow equation reads:
\eqn\nrgta{\Biggl[k{\partial\over{\partial k}}-{1\over 2}\bigl(d-2)
\bar\Phi{\partial\over{\partial\bar\Phi}}+d\Biggr]{\bar U}_k(\bar\Phi)
=-\int_{0}^{1}dt~
{z^d(t)~{\bar U}_k''(\bar\Phi)\over{ z^2(t)+(1-t){\bar U}''_k(\bar\Phi)}},}
where ${\bar U}_k(\bar\Phi)=\zeta^2k^{-d}U_k(\Phi)$,
$\bar\Phi=\zeta k^{-(d-2)/2}
\Phi$, $\zeta=\sqrt{2/S_d}
=\sqrt{(4\pi)^{d/2}\Gamma(d/2)}$, and the momentum scale $z(t)$
is given by Eqs. \pt, \zzt\ or \ptt\ for the three cases considered. 
That $S_d$ can be absorbed 
by a redefinition of $\bar\Phi$ is an indication of universality, i.e.,
critical exponents are independent of $S_d$, though the matrix 
elements in the linearized RG matrix are. 
In the above, the anomalous dimension
$\eta$ has been set to zero since we are only considering the flow of the 
potential.

To characterize the critical behavior of the theory, our first task is to
identify all the fixed points and then linearize  
RG about a particular fixed point. It is well known that
for the one-component scalar theory in $d=3$, there are only two fixed points:
one trivial Gaussian and one Wilson-Fisher, and the critical behavior is
dominated by the latter. No other continuum limit is known to exist.
This applies to both the sharp and the smooth cutoffs.
However, for a general smooth function $\rho_{k,\sigma}(p)$,
the location of the Wilson-Fisher fixed point
which will now depend on $\sigma$; 
at $\sigma=0$, however, it coincides with the Gaussian one.

At the fixed point(s), the theory exhibits scale invariance, i.e. 
$\partial_k {\bar U}_k^{*} = 0$, and the RG equation for the fixed
point potential becomes (dropping the subscript $k$):
\eqn\nrgfp{\eqalign{  -{1\over 2}\bigl(d-2\bigr)
\bar\Phi {\bar U}^{*'}(\bar\Phi)+d{\bar U}^{*}(\bar\Phi) &
 =-\int_{0}^{1}dt~
{z^d(t)~{\bar U}^{*''}(\bar\Phi)\over{ z^2(t)
+(1-t){\bar U}^{*''}(\bar\Phi)}} \cr
&
{{\rm s.}\atop\longrightarrow}~-{\rm ln}\Bigl[1+{\bar U}^{*''}
(\bar\Phi)\Bigr],}}
where the notation ${\rm s.}$ stands for the sharp cutoff limit. 
Analytically, the sharp limit has  
the following approximate non-trivial solution in the
large $\bar\Phi$ limit for $d=3$:
\eqn\fpv{ {\bar U}^{*}(\bar\Phi)=A{\bar\Phi}^6-{4\over 3}~{\rm ln}~
{\bar\Phi}-{2\over 9}-{1\over 3}~{\rm ln}(30A)-{1\over{150A{\bar\Phi}^4}}
+O({\bar\Phi}^{-6}),}
where $A$ is an arbitrary positive constant \morris. 
The RG flow about Eq. \fpv\ can be linearized by writing 
${\bar U}_k(\bar\Phi)={\bar U}^{*}(\bar\Phi)+{\bar v}(\bar\Phi)
e^{-\lambda{\rm ln} k}$, where ${\bar v}(\bar\Phi)$ obeys
\eqn\lirg{\eqalign{
-{1\over 2}(d-2)\bar\Phi{\bar v}'(\bar\Phi)-\bigl(\lambda-d)
{\bar v}(\bar\Phi) &= -\int_0^1dt~{z^d(t){\bar v}''
(\bar\Phi)\over{\bigl[z^2(t)+(1-t){\bar U}^{*''}(\bar\Phi)\bigr]^2}} \cr
&
{{\rm s.}\atop\longrightarrow}~
-{{{\bar v}''(\bar\Phi)}\over{ 1+{\bar U}^{*''}(\bar\Phi)}}.}}

We comment that the location of the fixed point
as well as ${\bar U}^{*}(\bar\Phi)$, being non-universal, naturally
all depend on the choice of the smearing function $\rho_k(p)$.  
In the one extreme where $\rho_{k,\sigma}(p)=\Theta(k-p)$,
linearlizing RG and retaining only the relevant operators yields
\eqn\fpp{ (\bar\mu^{2*},\bar\lambda^{*})=(0,0),~~
\Bigl(-{\epsilon\over{6+\epsilon}},{12\epsilon\over{B(6+\epsilon)^2}}\Bigr),}
where $B=1/{16\pi^2}$. 
However,
in the other extreme where the smearing function is taken to be a 
$k$-independent constant, $\rho_{k,\sigma}(p)=c$, the right-hand-side of
Eq. \nrgta\ then vanishes, and the RG flow simplifies to 
\eqn\nrd{ {dU_k(\Phi)\over dk}=0~~\Longrightarrow~~-{1\over 2}\bigl(d-2\bigr)
\bar\Phi {\bar U}^{*'}(\bar\Phi)+d{\bar U}^{*}(\bar\Phi)=0.}
The equation can be exactly solved to give ${\bar U}^{*}(\bar\Phi)
=A{\bar\Phi}^{2d/(d-2)}$. For $\epsilon=1$ or $d=3$, 
$U^{*}\sim {\bar\Phi}^6$. Thus, with vanishing contributions from quantum 
fluctuations, the only fixed point is Gaussian 
$(\bar\mu^{2*}, {\bar\lambda}^{*})=(0,0)$. However, this does not mean that
the theory is free, because we still have ${\bar U}^{*}(\bar\Phi)\ne 0$.

For the general case of a 
smooth cutoff function, however, it is rather difficult  
to obtain a non-truncated solution for Eq. \nrgfp, and we
circumvent the problem with polynomial truncation at order
$\phi^{2M}$. The remaining task is to 
solve a system of $M$ integro-differential 
equations for the coupling constants $g^{(2\ell)*}$, $\ell=1,\cdots, M$. 
We first determine the fixed-point solution analytically in the sharp
cutoff limit, so that we can then use this  
as an initial guess in our numerical root-finding subroutine for the
smooth case.  
The system of integro-differential equations is solved by first 
choosing a large value for the smoothness parameter $\sigma$
($1/\varepsilon$, $b$, or $m$), thereby making the smearing function 
rather sharp. 
The smoothness of the cutoff function is increased in small steps by 
decreasing the parameters, and the solution
of the previous step is used as the guess for finding the solution at the 
current step.  In this manner we are able to
track the solution associated with a physical fixed point from
the sharp cutoff limit to an
arbitrary smoothness.  

With the strategies outlined above, 
we now present the results for all three smearing functions considered
in Sec. II.

\medskip

\centerline{\epsfbox{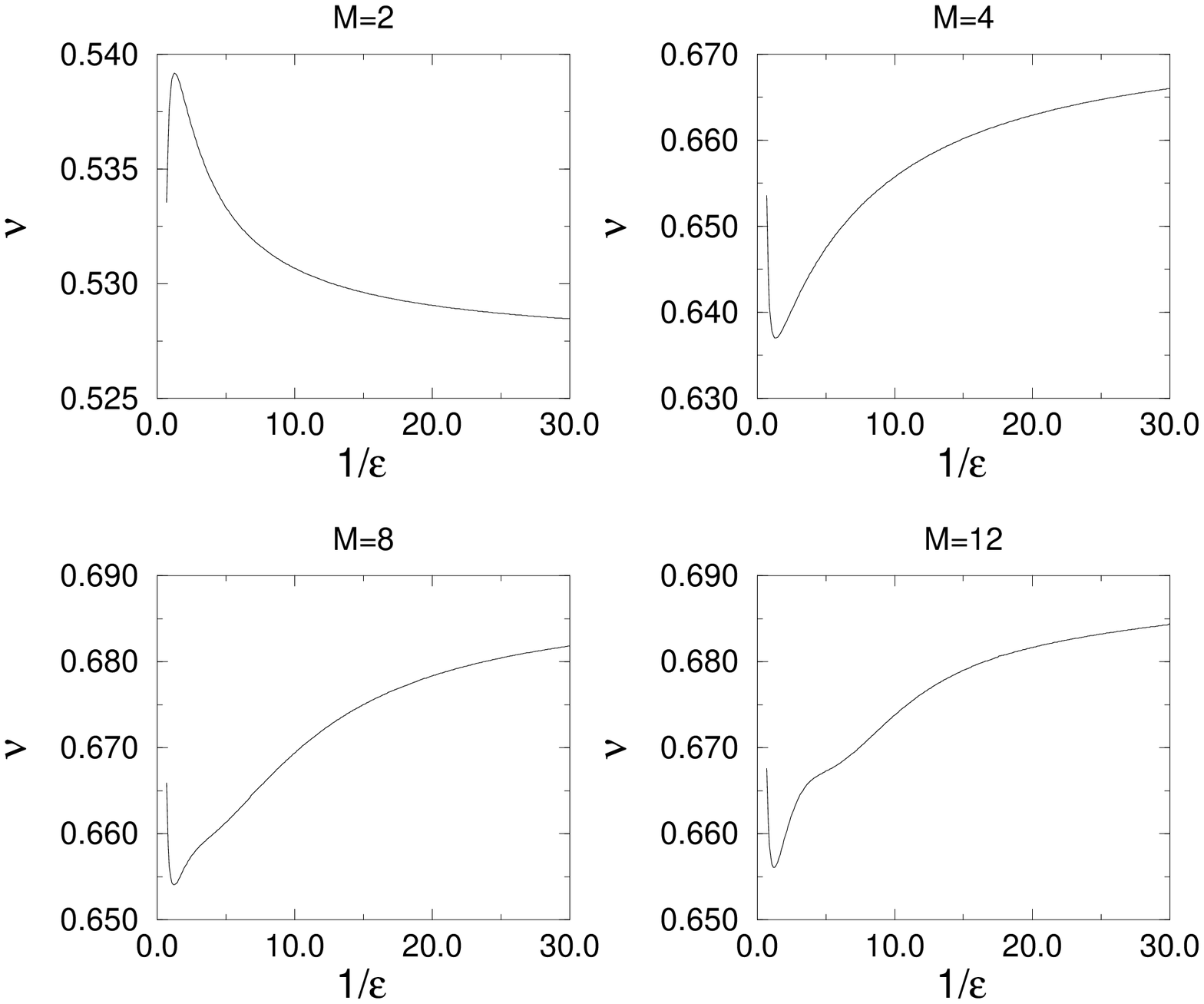}}
\medskip
{\narrower
{\sevenrm
{\baselineskip=8pt
\itemitem{Figure 1.}
Critical exponent $\scriptstyle \nu$ as a function of 
$\scriptstyle 1/\varepsilon$ using the hyperbolic
tangent smearing function.  Results for four different levels of polynomial
truncation are shown.
\bigskip
}}} 

In Fig. 1 we plot the dependence of $\nu$ in $d=3$ as a function
of the (inverse) smoothness parameter $\varepsilon^{-1}$ 
at four different levels of truncation of $U^{*}(\Phi)$.
From the figure, we see that
$\nu$ varies by 2-5\% over the range shown.  
The variation would become even larger had we taken $\varepsilon^{-1}\to 0$,
where the smearing function becomes very flat and
there is practically no blocking. The exponents in this limit are 
those obtained in the mean-field approximation, e.g.,
$\nu=0.5$.
It is also interesting to note that at each order in $M$ there
is a dip, or an extremum whose position  
changes only slightly with $M$. The non-monotonic behavior 
shows that when going to the sharp limit by gradually increasing
$\varepsilon^{-1}$, $\nu$ first deviates from the sharp result and then
begins to converge when $\rho_{k,\varepsilon}(p)$ becomes sufficiently sharp. 

When an exponential smearing function is employed instead, we observe a
similar behavior for $\nu$, as depicted in Figure 2. Once more, the value
varies by 2-5\% and an extremum is found for each $M$.  
The qualitative feature remains the same 
for a power-law smearing function, as demonstrated in Figure 3.
However, in this case the variation of $\nu$ is only about 2-3\%. 

\medskip

\centerline{\epsfbox{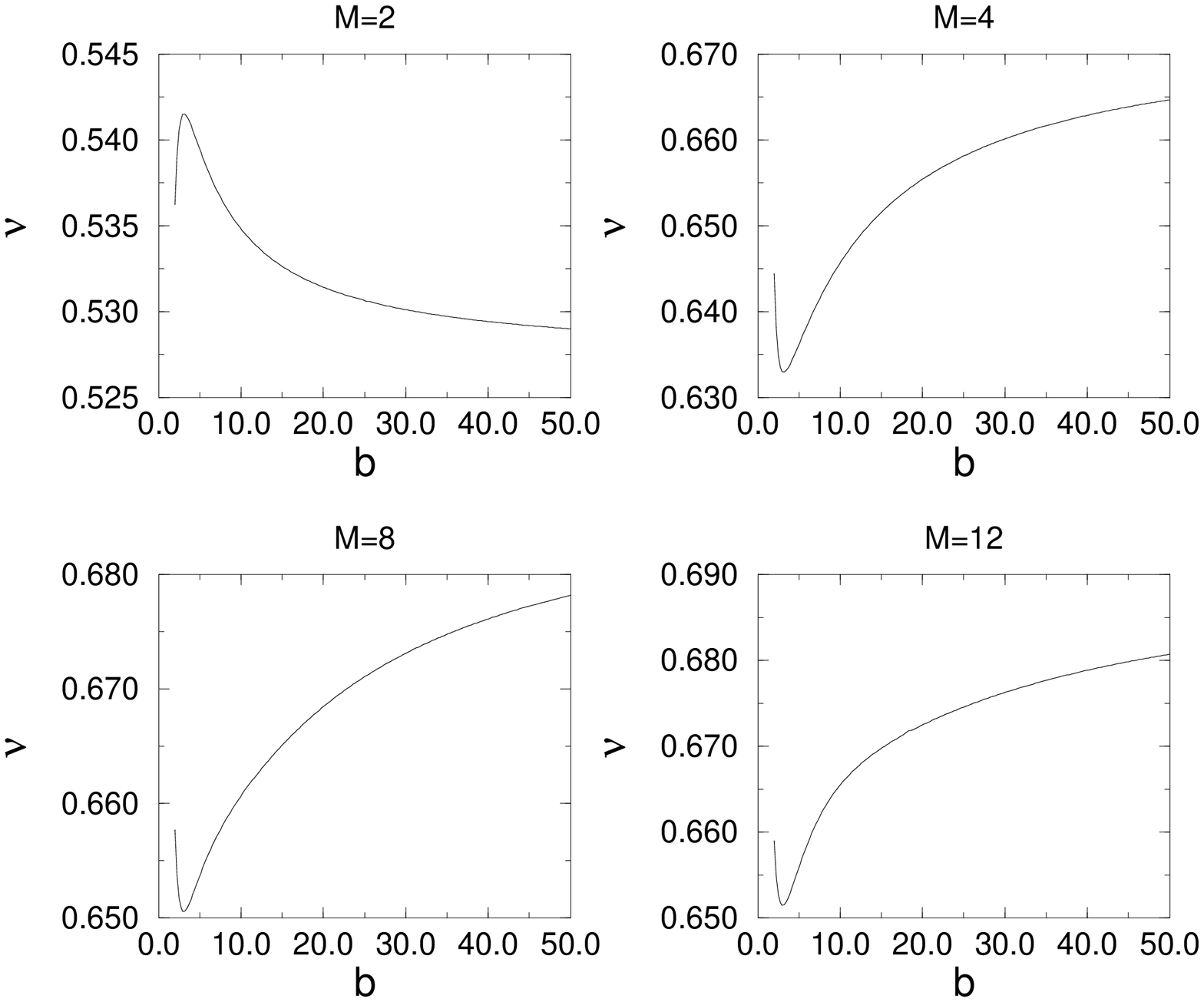}}
\medskip
{\narrower
{\sevenrm
{\baselineskip=8pt
\itemitem{Figure 2.}
Critical exponent $\scriptstyle \nu$ as a function of 
$\scriptstyle b$ for the exponential
smearing function.  Results for four different levels of polynomial
truncation are shown.
\bigskip
}}} 

\medskip

\centerline{\epsfbox{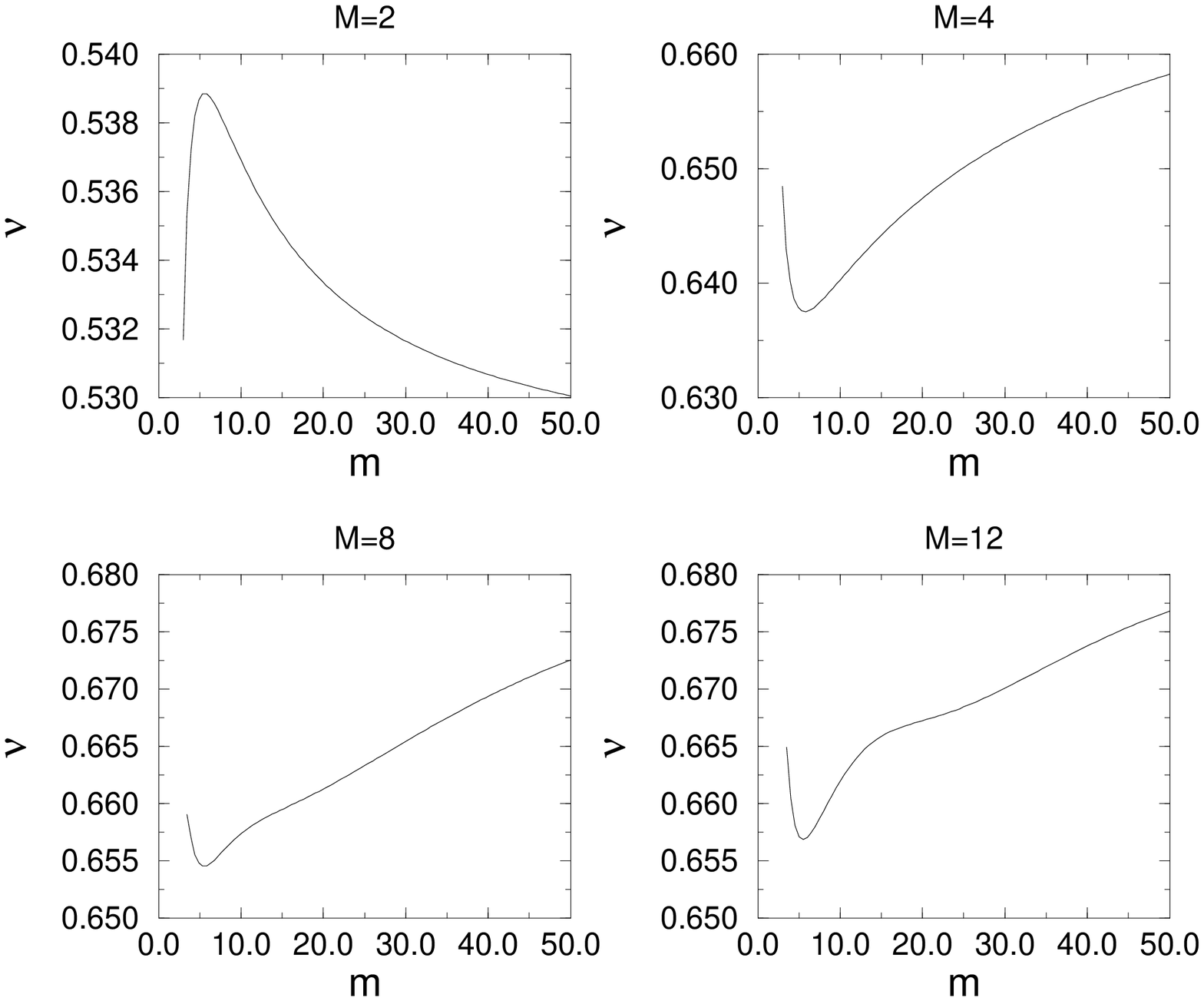}}
\medskip
{\narrower
{\sevenrm
{\baselineskip=8pt
\itemitem{Figure 3.}
Critical exponent $\scriptstyle \nu$ as a function of 
$\scriptstyle m$ for the power law
smearing function.  Results for four different levels of polynomial
truncation are shown.
\bigskip
}}} 

\centerline{\bf IV. SCHEME DEPENDENCE}
\medskip
\nobreak
\xdef\secsym{4.}\global\meqno = 1
\medskip
\nobreak

Even though non-truncated solutions for 
the fixed-point potential ${\bar U}^{*}(\bar\Phi)$ can be obtained in special 
cases \morris, this is not possible in general. Any approximation
inevitably brings in scheme-dependent effects. Our results are seen to
depend both on the order of truncation $M$
as well as the smoothness parameter $\sigma$. We discuss below the 
implications of such dependence.
\medskip

\centerline{\bf A. Truncation Dependence} 
\medskip

At order $M$ we obtain $M$ fixed-point solutions parameterized by
the critical coupling constants ($g^{(2)*}$ $\cdots$ $g^{(2M)*}$),  
each having its own eigenvectors and eigenvalues which in turn
can be used for calculating the critical exponents.  
However, only one of the solutions corresponds to 
the Wilson-Fisher fixed point which is characterized by one relevant
operator, with the rest being irrelevant. The trivial Gaussian solution
with all $g^{(2\ell)*}=0$ is also obtained but is of no interest to us.
Computational artifacts are induced with a polynomial truncation 
of the blocked potential 
${\bar U}_k(\bar\Phi)$ \aoki \morris \mike.  
For example, as $M$ is increased, numerous unphysical fixed points are
also generated. 

Thus, to isolate the physically meaningful solution from the 
unphysical ones in the smooth parameterization, what we have done numerically
was to first analyze the sharp cutoff limit
of Eq. \nrgfp\ and identify the solution associated with 
the Wilson-Fisher fixed point. The salient feature of this solution is that
it is stable against an increase in $M$. In other words, the value of
$g^{(2\ell)*}$ is not sensitive to whether higher order coupling constants, 
e.g., $g^{(2\ell+2)*}$, are included or not. On the other hand,
unphysical solutions are sensitive to the higher order equations and 
their eigenvalues will fluctuate in an unpredictable manner \margaritis.  

Another prominent artifact of the polynomial truncation is the oscillation of
the critical exponents with $M$ when a sharp cutoff is used.
The values of the critical coupling constants also alternate in sign.
The oscillations are due to the nonanalyticity of the 
critical blocked potential ${\bar U}^{*}(\bar\Phi)$
and can be understood from the fact
that an expansion of the right-hand-side of the sharp cutoff flow equation
${\rm ln}(1+{{\bar U}^{*''}(\bar\Phi)})
=\sum_{\ell=1}^{\infty}(-1)^{(\ell+1)}({\bar U}^{*''}(\bar\Phi))^{\ell}
/{\ell}$ also yields a power series of $\bar\Phi$ with alternating sign.
Even though the convergence has been shown to 
improve significantly by expanding 
${\bar U}_k(\bar\Phi)$ around its moving minimum, there still remains
residual scheme dependence \aoki.  In addition, an expansion about
the minimum fails in the broken phase of $O(N)$ as stated in the Introduction.
Therefore, it is desirable to have a numerical technique which allows for an 
efficient determination of the critical properties, and remove  
or reduce the aforementioned spurious effects.

Another subtle issue remains in this approximation is:
why can we not reproduce the exact
critical exponent(s) by keeping only the renormalizable coupling constants
in our computation and in the identification of the fixed point? The
main reason lies in that fact 
the fixed point of the truncated solution is 
applicable only up to the
operators which have been neglected. At the 
point which we call a ``fixed point,'' 
the would-be ignored set of irrelevant operators does not vanish but
continues to evolve.
On the other hand, in the exact RG approach, the fixed-point
condition implies a complicated cancellation between different operators.
When all the operators are present, the cancellation is complete and 
we should be able to eliminate the $\rho_{k,\sigma}(p)$-dependence 
in the scaling laws around the fixed point as well as in the
expression for the critical exponents.

\medskip
\medskip

\centerline{\bf B. Smoothness Dependence} 
\medskip

In the alternative Wilson or Polchinski RG approach, one can show
that when
the effective action is expanded in terms of derivatives to order
$n$, only $2n$ parameters are required to absorb the scheme dependence \ball. 
Therefore, at the leading order of $U_k(\Phi)$ 
with $n=0$, the critical exponents must be
scheme-independent, i.e., the same results are obtained regardless of
whether the smearing function is sharp, exponential or power-like. 
In fact, $\rho_{k,\sigma}(p)$ can be conveniently
absorbed by a suitable redefinition of the field variable \ref\commellas.
Scheme depedence appears, however, when the wavefunction renormalization
$Z_k(\Phi)$ is taken into consideration at the next order. 

On the other hand, Eq. \nrgta\ derived with LPA 
does not possess this scaling property, and the functional
form of the flow equation is explicitly dependent of $\rho_{k,\sigma}(p)$.
The dependence, nevertheless, is expected to be
negligible in the calculation of physical quantities such as the critical 
exponents. This universality hypothesis, however,
must be substantiated by solving the 
integro-differential equation Eq. \nrgta\ exactly without polynomial
expansion. 

From Figures 1, 2 and 3, the critical exponent $\nu$ clearly depends 
on the smoothness
parameter $\sigma$ ($\varepsilon^{-1}$, $b$ and $m$). This comes as a 
consequence of polynomial truncation. That is, because of
the polynomial truncation scheme employed, physical results now depend both
on the level of truncation $M$ as well as on $\sigma$. 

How can such dependence be reconciled with universality? The key to the 
problem here again lies in the role played by the irrelevant operators 
in the approximate solution. 
If we solve the RG equation for the full theory and come
down to the IR regime, the dependence on the initial condition for
the irrelevant operators must die out according to the universality
hypothesis. The explicit verification of this scenario by reaching 
the IR fixed point in principle requires an infinite number of iterations. 
This is not what was followed in this work. Instead we move close to the
UV fixed point and a linearized RG prescription 
is utilized to deduce the critical exponents and the corresponding
scaling laws. Using the derivative expansion and the
polynomial truncation schemes gives the advantage of 
a readily accessible (approximate) fixed-point solution; 
however, we are no longer certain what the omitted irrelevant operators 
do. If they are not scale invariant at the fixed point - the 
likely result of the truncation, then their variation amounts 
to the change of the overall scale factor of the exact solution
of the theory. This is because the irrelevant coupling
constant set of the theory influences an overall scale factor. 
In order to minimize this scale dependence the truncation scheme
must be ``improved''.

The comparison with the exact RG offers a direction
of improvement in our approximation:
to minimize the number and the strength of the neglected irrelevant operators.
As far as the $\rho_{k,\sigma}(p)$-dependence is concerned,
the momentum dependent, i.e. higher order derivative terms come
into consideration which are quadratic in the field $\Phi$. Therefore, we 
aim to minimize the number of such terms generated in the RG
flow. The cutoff-function $\rho_{k,\varepsilon}(p)$ is strongly momentum 
dependent in an interval $\Delta p=\varepsilon k$. According to the uncertanity 
principle this yields non-local interactions on the length scale 
$\Delta x=1/\varepsilon k$. In order to eliminate such a higher order
derivative effect from the theory we 
must look for a cutoff which is as smooth as
possible in the momentum space. This would yield a better convergence when
the irrelevant derivative operators are truncated, in accord with the 
spirit of LPA. 

However, there are also higer-order field operators and 
we should keep $M$ as low as possible. 
We emphasize that the purpose of employing a truncated 
polynomial expansion of $U_k(\Phi)$ is to devise a simple algorithm which 
allows us to reproduce the nonperturbative features, particularly
the fixed-point structure embedded in LPA, or even the
blocked action ${\widetilde S}_k[\Phi]$, without having to solve the full
flow equation for $U_k(\Phi)$. Thus, to go to a large $M$ not only is 
computationally laborious but also in conflict with the original 
goal. 

In Fig. 4 the truncation dependence of $\nu$ obtained
using the exponential smearing function given in Eq. \bsm\ is depicted.
The sharp cutoff results are also included for comparative purpose.
We notice the general trend of a more pronounced oscillatory behavior 
as $b$ is increased toward the sharp limit. This again show
that the sharp cutoff
is not the {\it best} candidate if we look for the smallest set of 
operators which
yields the most rapid convergence for the physical observables.
Below we demonstrate how this can be achieved with an optimization scheme.

\medskip
\medskip

\centerline{\epsfbox{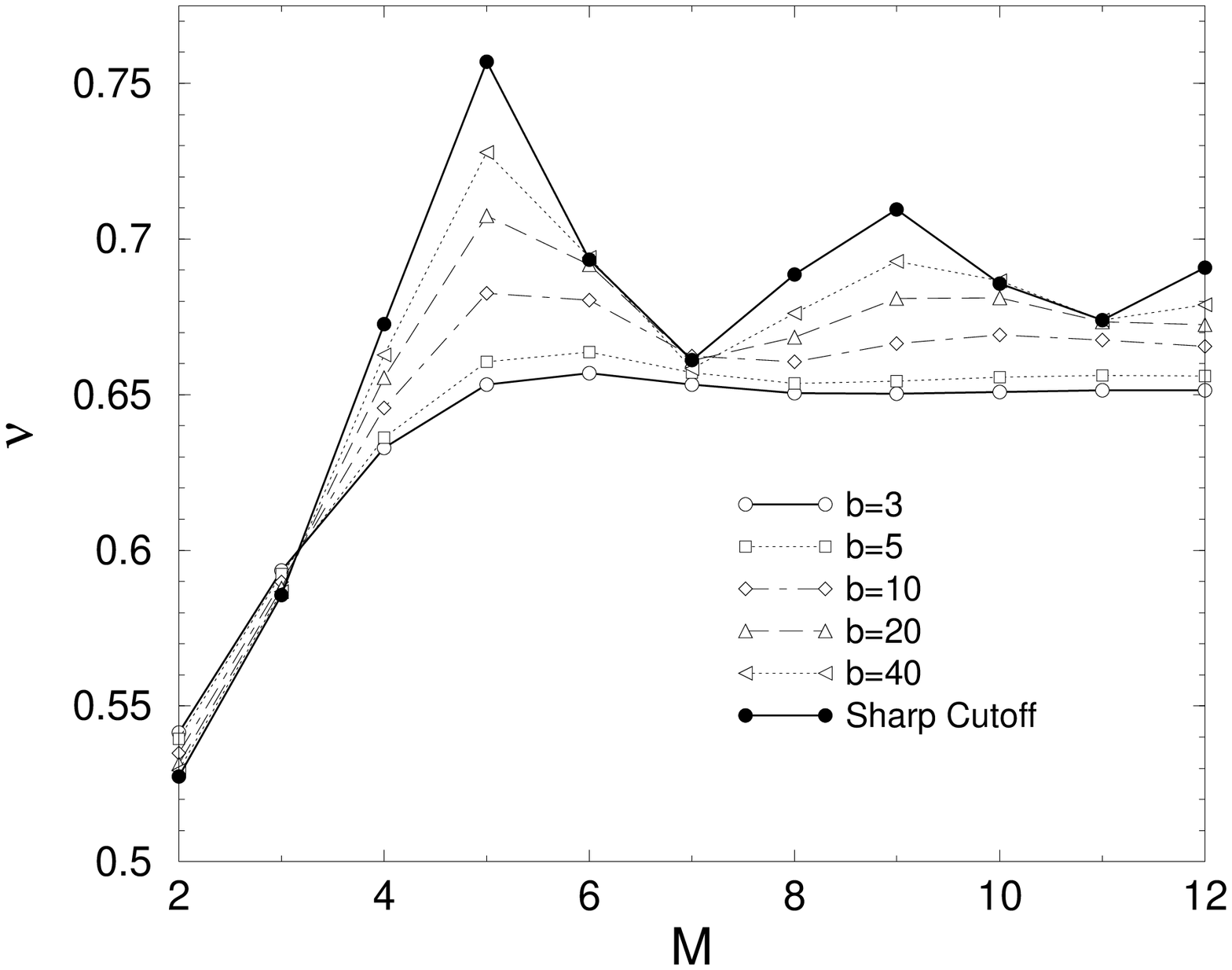}}
\medskip
{\narrower
{\sevenrm
{\baselineskip=8pt
\itemitem{Figure 4.}
Critical exponent $\scriptstyle \nu$ as a function of $\scriptstyle M$ 
using the exponential smearing function. Results for several values of 
$\scriptstyle b$ are shown for comparison.  
\bigskip
}}} 

\medskip
\medskip
\centerline{\bf C. Optimization}
\medskip
\medskip

For a sharp cutoff, with a clear boundary between the high and 
the low modes, the IR cutoff is determined unambiguously as $k$. 
However, we wish to choose the cutoff to be as smooth 
as possible in order to minimize the generation of non-local, 
higher-derivative irrelevant interactions, as well as the order of 
truncation $M$. 
A general smooth smearing function $\rho_{k,\sigma}(p)$ results in a shift in
the actual value of the IR cutoff, or the location of the peak of 
$-\partial\rho_{k,\sigma}(p)/\partial p$.  When $\rho_{k,\sigma}(p)$  
is too smooth, we no longer have a well-defined cutoff, i.e., the relation 
between the parameter $k$ and the momenta of the modes appearing in the
evolution equation. 

As we have shown before, with the truncation scheme employed, the 
fixed-point solutions will vary with $M$ and the    
would-be irrelevant operators will continue to
contribute and evolve around the these approximate solutions. Thus,
we would not expect {\it a priori} an accurate result for
the critical exponents in this case.
Nevertheless, we notice that
for each of the three smearing functions discussed in Sec. II, there
is an ``optimal'' value of $\sigma$ for
which $\nu$ converges most rapidly. For example, the optimal value for 
the exponential function is found to be $b=3$, which
incidentally coincides with the value of the extremum shown in Figure 2.  
When all three smearing functions are superimposed, it can be
seen that they have approximately the same shape or smoothness, as
illustrated in Fig. 5. That is, the most optimal smoothness
does not depend on the detailed form
of the smearing, i.e., hyperbolic tangent, exponential or 
power-law.

\medskip
\medskip

\centerline{\epsfbox{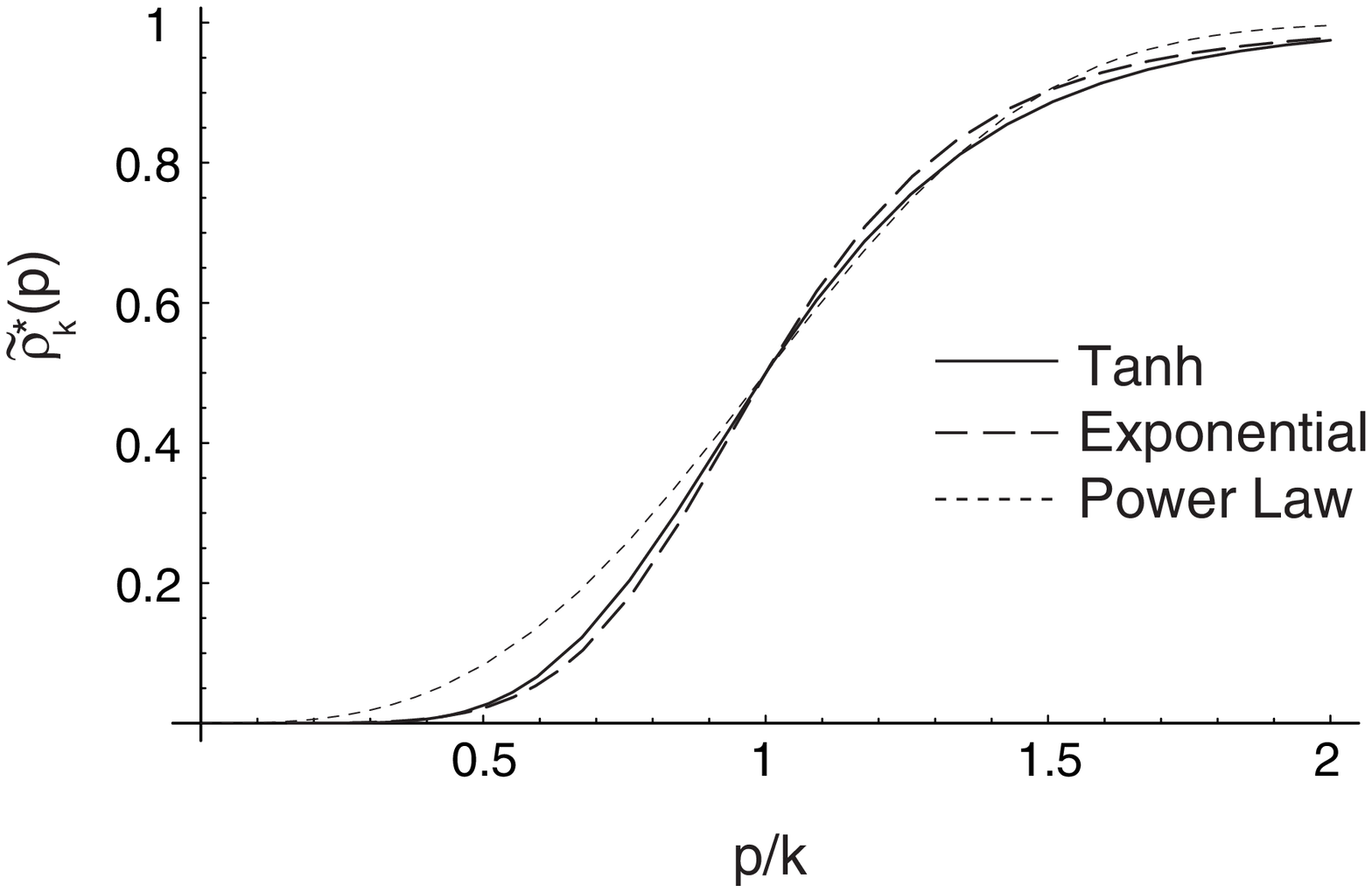}}
\medskip
{\narrower
{\sevenrm
{\baselineskip=8pt
\itemitem{Figure 5.}
Comparison of the three optimized smearing functions. 
\bigskip
}}} 

The truncation dependence of $\nu$
for the three optimized smearing functions is ilustrated in Fig. 6.
Although there remains small oscillations, 
we see a dramatic improvement in the convergence. 
In fact, the variation beyond $M=7$ is only 1-2\%.
Our optimized value gives $\nu=0.65(5)$ which is closer to the 
world's best value $0.631(2)$ \ref\zinn\ than $0.68(9)$ obtained by Aoki
{\it et. al.} using an expansion around the moving minimum.  
We also remark that $M$, the number of operators involved in our calculation,
is considerably smaller for a reasonably accurate estimate 
of $\nu$ \aoki.  However, by truncating the potential at $M=12$,
we have not been able to observe
the oscillation of $\nu$ with $M$ at a four-fold periodicity; an unambiguous
observation of such behavior generally would require $M \ge 20$  
\aoki\morris.

\medskip
\medskip

\centerline{\epsfbox{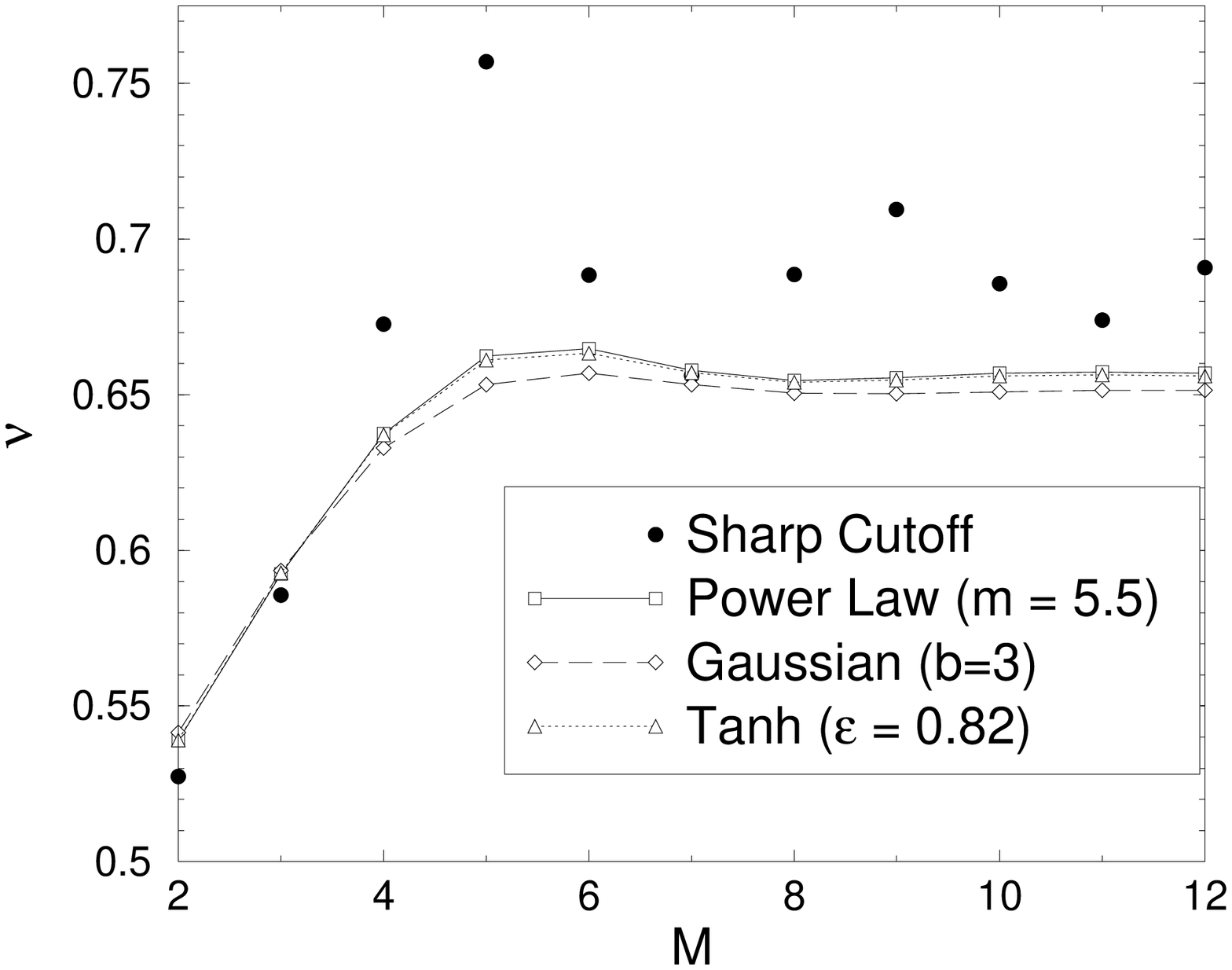}}
\medskip
{\narrower
{\sevenrm
{\baselineskip=8pt
\itemitem{Figure 6.}
Critical exponent $\scriptstyle \nu$ as a function of the level of 
polynomial truncation
for all three optimized smearing functions and the sharp cutoff. 
\bigskip
}}} 

Why is $\sigma$ related to the convergence of the series expansion?
Physically, the parameter dictates the manner in which the quantum 
fluctuations modes are integrated over. When $\sigma$ is very small and the 
corresponding smearing function is too smooth, certain fast-fluctuating 
modes are only damped but not integrated over due to the lack of a well-defined
boundary. In addition, some slowly-varying degrees of freedom which should
be kept are integrated.
As a result of the ``mistreatment'' of both the fast and the slow modes,
the RG trajectory is greatly distorted, and the would-be
irrelevant higher-order operators continue to evolve near the critical
point obtained at order $M$. 

The complicated interplay between the irrelevant operators can also be 
seen from the non-monotonic
dependence of $\nu$ on $\sigma$ observed in Figures 1 to 3.
An extremum is found at $\sigma_c$ for all orders of $M$. While
for $\sigma_c$ is a local maximum for $M=2$, it becomes a local minimum
for $M \ge 3$. The flip is clearly due to the inclusion of the irrelevant
(marginal by power counting) $\Phi^6$ operator.

In general, as $\sigma$ increases and more fast
modes are being included in the loop integrations, more cancellations
between the effects of the irrelevant operators take place and 
the theory moves closer to
the {\it true} Wilson-Fisher fixed point found in the exact approach.  
Nevertheless, when $\sigma$ becomes too large and the resulting smearing
function is too sharp, non-local effects begins to set in and the theory
drifts away from the true RG trajectory. At $\sigma=\infty$ where 
the smearing function becomes a step function $\Theta(k-p)$, non-local 
effects become maximal and LPA is no longer adequate. 
One must then include derivative 
operators to all orders in the blocked action ${\widetilde S}_k[\Phi]$ 
in order to arrive at a scheme-independent result.

Thus, we see that $\sigma$ monitors the manner in which the irrelevant 
operators contribute to the flow of the theory. When the most optimal
smoothness is reached, the maximum cancellation between these operators 
takes place, leading to the fastest convergence in the polynomial truncation.

In Sec. II we have also discussed an alternative  
mean approach, which is based on taking the sharp cutoff limit 
followed by a substitution of $\theta_0=1/2$ to avoid ambiguity, and 
yields Eq. \smrf\ as the RG flow. We solve the equation and 
depict the result for $\nu$ in Figure 7. 

Contrary to the sharp 
approach which results in an oscillatory behavior, $\nu$ is seen to converge 
rapidly. Such a change in the convergence is 
intimately related to 
the structure of the untruncated fixed-point potential
${\bar U}^{*}(\bar\Phi)$ \tmorris\ which now obeys
\eqn\mfp{ -{1\over 2}\bigl(d-2\bigr)
\bar\Phi {\bar U}^{*'}(\bar\Phi)+d{\bar U}^{*}(\bar\Phi)
=-{2{\bar U}^{*''}(\bar\Phi)\over{ 2+ {\bar U}^{*''}(\bar\Phi)}}.}

\medskip
\medskip

\centerline{\epsfbox{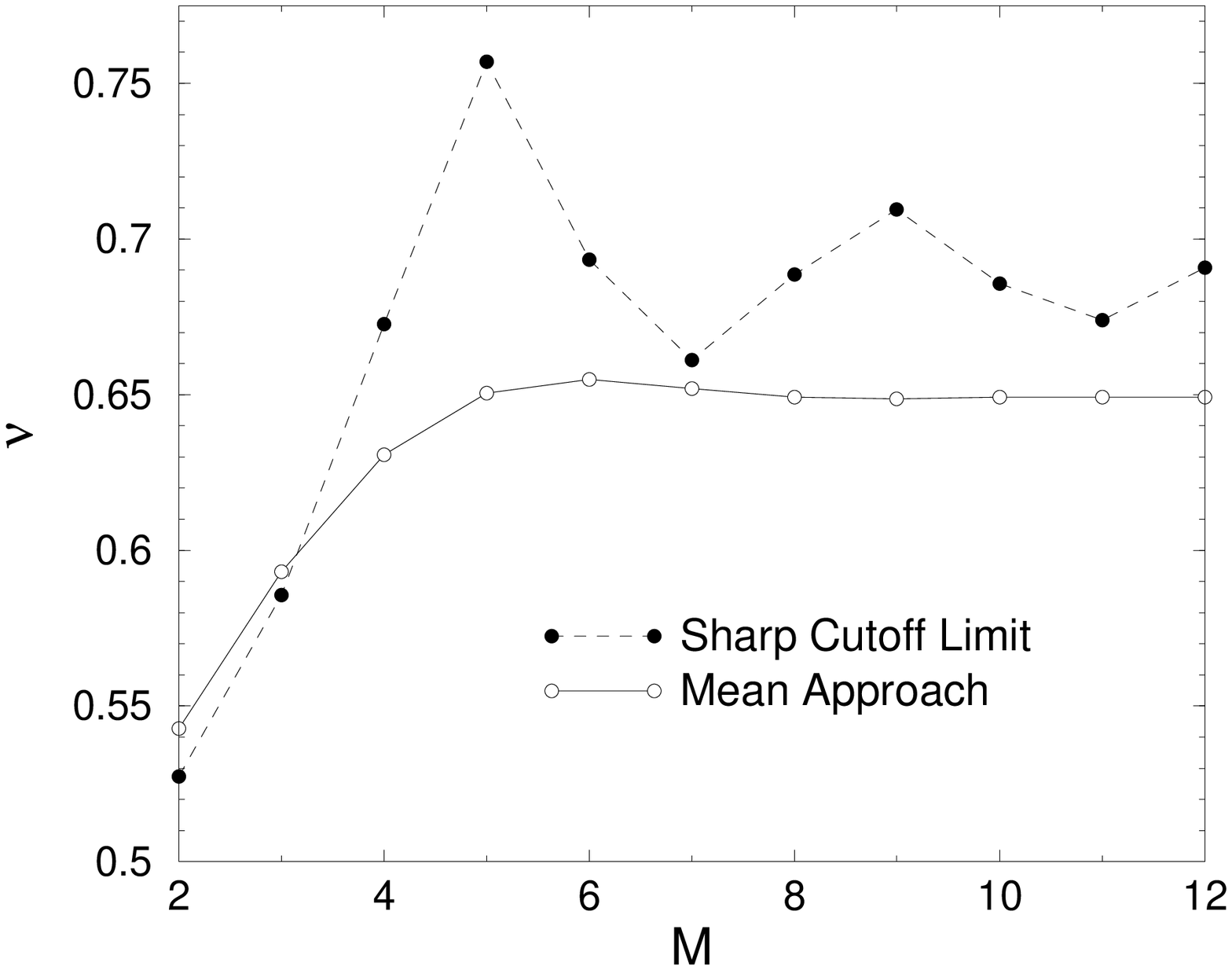}}
\medskip
{\narrower
{\sevenrm
{\baselineskip=8pt
\itemitem{Figure 7.}
Critical exponent $\scriptstyle \nu$ as a function of the level of 
polynomial truncation
obtained with the mean approach and the sharp cutoff. 
\bigskip
}}} 

Even though Eq. \mfp\ still cannot be solved analytically, its 
improvement in convergence can be shown by examining
the right-hand-side of the equation representing the quantum corrections. 
By making a Taylor expansion of the expression with respect to 
$\bar U_k''(\bar\Phi)$,
we see that it has a radius of convergence $2/{\bar U}_k''$
which is twice that of ${\rm ln}[1+{\bar U}_k''(\bar\Phi)]$ in Eq. \nrgfp.
Since a general smooth cutoff always has a greater convergence radius, 
the resulting critical exponents will converge more rapidly compared with
the sharp in the polynomial truncation scheme. To describe 
the amplitude and the periodicity of the oscillation more quantitatively, 
however, would require a detailed knowledge of the singularities of 
the untruncated potential ${\bar U}^{*}(\bar\Phi)$.

\medskip
\medskip
\centerline{\bf V. SUMMARY AND DISCUSSIONS}
\medskip
\nobreak
\xdef\secsym{5.}\global\meqno = 1
\medskip
\nobreak

In the present work we have demonstrated how one can improve the 
convergence of critical exponents calculated using polynomial truncations
of the RG flow equation for the blocked potential $U_k(\Phi)$.
Since in the sharp limit, $\nu$ oscillates with $M$, the order of
truncation, we examined three parameterizations
of a smooth cutoff function, hyperbolic tangent, exponential, and power-law,
and made an attempt to eliminate as much as possible such
unphysical artifacts.
We find that there exists an optimal smoothness value which gives
the most rapid convergence as a function of $M$. This is
due to the maximal cancellation of the effects generated by the
irrelevant operators in the blocked action.  
Our optimal smearing functions yield
$\nu = 0.65(5)$ for $M \ge 7$ with a variation of 1-2\% 
between the three cases. Had $\sigma$ been too large, we would have to
incorporate the higher-order derivative operators to account for the 
non-local effects generated in the course of RG evolution.

We have also learned from comparing the results for the sharp limit 
and the mean approach (depicted in Figure 7) that the 
convergence behavior of the polynomial truncation scheme 
is intimately related to the non-truncated solution of ${\bar U}_k(\bar\Phi)$. 
The mean approach results in a larger radius of convergence, and hence
a more rapid convergence. It remains to see if the same ``trick'' can be
utilized for more complicated systems as well. 

In light of the success of our optimized RG prescription, we can readily
extend the formalism to address other issues such as the $O(N)$ models, 
the spinodal instability \ref\spino, 
gauge theories or chiral symmetry breaking. 
It would also be interesting to compare our usual momentum-cutoff 
approch with an alternative internal-space RG, which is a
functional generalization of the Callan-Symanzyk equation,
based on an infinitesimal variation of an internal-space 
parameter such as the mass scale \ref\intsp.

\bigskip
\goodbreak
\bigskip
\centerline{\bf ACKNOWLEDGMENTS}
\medskip
\nobreak
S.-B. L. is grateful to Y.-C. Tsai for valuable discussions. 
M.S. would like to thank J.O. Andersen and E. Braaten.
This work is supported in part by funds provided by the National Science 
Council of Taiwan under contract \#NSC-88-2112-M-194-006, and by 
the National Science Foundation under Grant No. PHY-9800964.

\bigskip
\medskip
\medskip
\centerline{\bf REFERENCES}
\medskip
\medskip
\nobreak

\item{\wilson} K. Wilson, {\it Phys. Rev.} {\bf B4} (1971) 3174;
K. Wilson and J. Kogut, {\it Phys. Rep.} {\bf 12C} (1975) 75.
\medskip
\item{\polonyi} See, for example, J. Polonyi, {\it Physics of the 
Quark-Gluon Plasma}, hep-ph/9509334.
\medskip
\item{\lp} S.-B. Liao, J. Polonyi and D. P. Xu, {\it Phys. Rev.}
{\bf D51} (1995) 748.
\medskip
\item{\sb} S.-B. Liao and J. Polonyi, {\it Ann. Phys.} {\bf 222} (1993) 122
and {\it Phys. Rev.} {\bf D51} (1995) 4474.
\medskip
\item{\tmorris} T. R. Morris, {\it Int. J. Mod. Phys.} {\bf A9} (1994) 2411; 
{\it Nucl. Phys.} {\bf B509} (1998) 637, and 
references therein.
\medskip
\item{\wegner} F.J. Wegner and A. Houghton, {\it Phys. Rev} 
{\bf A8} (1972) 401.
\medskip
\item{\hasenfratz}
A. Hasenfratz and P. Hasenfratz, {\it Nucl.} 
{\bf B270} (1986) 687. 
\medskip
\item{\margaritis}
A. Margaritis, G. Odor and A. Patkos, {\it Z. Phys.} {\bf C39} (1988) 109.
\medskip
\item{\aoki} K. Aoki, K. Morikawa, W. Souma, J. Sumi and H. Terao,
{\it Prog. Theor. Phys.} {\bf 95} (1996) 409 and {\bf 99} (1998) 451.
\medskip 
\item{\polchinski} J. Polchinski, {\it Nucl. Phys.} 
{\bf B231} (1984) 269.
\medskip
\item{\others} 
B. Bergerhoff, {\it Phys. Lett.} {\bf B437} (1998) 381;
\medskip
A. Bonanno, V. Branchina, H. Mohrbach and D. Zappala, hep-th/9903173;
\medskip
J. Shafer and J. Shepard, {\it Phys. Rev.} {\bf D55} (1997) 4990;
\medskip
D. Dalvit and F. Mazzitelli, {\it Phys. Rev.} {\bf D54} (1996) 6338;
\medskip 
J. Rudnick, W. Lay and D. Jasnow, {\it Phys. Rev.} {\bf E58} (1998) 2902;
\medskip
N. Tetradis and C. Wetterich, {\it Nucl. Phys.} 
{\bf B383} (1992) 197 and {\bf B422} (1994) 541; 
\medskip
K. Ogure and J. Sato, {\it Phys. Rev.} {\bf D57} (1998) 7460;
\medskip
M. D'Attanasio and M. Pietroni,  {\it Nucl. Phys.} {\bf B472} (1996) 711; 
\medskip
M. Bonini, M. D'Attanasio and G. Marchesini,
{\it Nucl. Phys.} {\bf B437} (1995) 163;
\medskip
\item{\morris} T. R. Morris,
{\it Phys. Lett.} {\bf B334} (1994) 355 and {\bf B329} (1994) 241;
\medskip
\item{\nicoll} J. F. Nicoll, T. S. Chang and H. E. Stanley, {\it Phys. Rev.
Lett.} {\bf 32} (1974) 1446, {\bf 33} (1974) 540; {\it Phys. Rev.}
{\bf A13} (1976) 1251.
\medskip
\item{\jens} Jens O. Andersen and M. Strickland, cond-mat/9811096.
\medskip
\item{\ball} R. D. Ball, P. E. Haagensen, J. I. Latorre and E. Moreno,
{\it Phys. Lett.} {\bf B347} (1995) 80.
\medskip
\item{\mike} S.-B. Liao and M. Strickland, {\it Nucl. Phys.} {\bf B497}
(1997) 611 and {\it Phys. Rev.}
{\bf D52} (1995) 3653.
\medskip
\item{\sbl} S.-B. Liao, {\it Phys. Rev.} {\bf D53} (1996) 2020, and
{\it ibid.} {\bf D56} (1997) 5008.
\medskip
\item{\bergerhoff} 
B. Bergerhoff and J. Reingruber, hep-th/9809251.
\medskip
\item{\cons} S.-B. Liao and M. Strickland, {\it Nucl. Phys.} {\bf B532}
(1998) 753.
\medskip
\item{\commellas}
J. Comellas and A. Travesset, {\it Nucl. Phys.} {\bf B498} (1997) 539;
J. Commellas, {\it ibid.} {\bf B509} (1998) 662.
\medskip
\item{\zinn} J. Zinn-Justin, {\it Quantum Field Theory and Critical
Phenomena}, 3rd ed. (Clarendon, Oxford, 1996).
\medskip
\item{\spino} J. Alexandre, V. Branchina and J. Polonyi, {\it Phys. Rev.}
{\bf D58} (1998) 016002.
\medskip
\item{\intsp} J. Alexandre and J. Polonyi, {\it Renormalization group 
for the internal space}, submitted to {\it Phys. Rev.}, hep-th/9902144.
\vfill
\eject
\end